\documentclass[showpacs, preprintnumbers, pra, superscriptaddress, floatfix, twocolumn, longbibliography, footinbib]{revtex4-1}
\usepackage{natbib}
\usepackage{graphicx}
\usepackage{bm}
\usepackage{hyperref}
\usepackage{color}

\usepackage{amssymb}
\usepackage{amsmath}
\usepackage[applemac]{inputenc}

\newcommand{\kn}{{n\mathbf{k}}}

\newcommand{\pd}{{\phantom\dag}}

\newcommand{\jac}{Pt$_2$HgSe$_3$}
\newcommand{\jacpd}{Pd$_2$HgSe$_3$}

\begin{document}
\title{Dual topology in Jacutingaite \jac}

\author{Jorge I. Facio}
\affiliation{IFW Dresden and W{\"u}rzburg-Dresden Cluster of Excellence ct.qmat, Helmholtzstr. 20, 01069 Dresden, Germany}
\author{Sanjib Kumar Das}
\affiliation{IFW Dresden and W{\"u}rzburg-Dresden Cluster of Excellence ct.qmat, Helmholtzstr. 20, 01069 Dresden, Germany}

\author{Yang Zhang}
\affiliation{IFW Dresden and W{\"u}rzburg-Dresden Cluster of Excellence ct.qmat, Helmholtzstr. 20, 01069 Dresden, Germany}
\affiliation{Max Planck Institute for the Physics of Complex Systems, N\"{o}thnitzerstr.~38, 01187 Dresden, Germany}

\author{Klaus Koepernik}
\affiliation{IFW Dresden and W{\"u}rzburg-Dresden Cluster of Excellence ct.qmat, Helmholtzstr. 20, 01069 Dresden, Germany}

\author{Jeroen van den Brink}
\affiliation{IFW Dresden and W{\"u}rzburg-Dresden Cluster of Excellence ct.qmat, Helmholtzstr. 20, 01069 Dresden, Germany}
\affiliation{Department of Physics, Technical University Dresden and W{\"u}rzburg-Dresden Cluster of Excellence ct.qmat, Helmholtzstr.~10, 01062 Dresden, Germany}
\affiliation{Department of Physics, Washington University, St. Louis, MO 63130, USA}

\author{Ion Cosma Fulga}
\affiliation{IFW Dresden and W{\"u}rzburg-Dresden Cluster of Excellence ct.qmat, Helmholtzstr. 20, 01069 Dresden, Germany}

\begin{abstract}
Topological phases of electronic systems often coexist in a material, well-known examples being systems which are both strong and weak topological insulators. 
More recently, a number of materials have been found to have the topological structure of both a weak topological phase and a mirror-protected topological crystalline phase.
In this work, we first focus on the naturally occurring mineral called Jacutingaite, \jac, and show based on density-functional calculations that it realizes this dual topological phase and that the same conclusion holds for \jacpd.
Second, we introduce tight-binding models that capture the essential topological properties of this dual topological phase in materials with three-fold rotation symmetry and use these models to describe the main features of the surface spectral density of different materials in the class.

\end{abstract}

\maketitle

\section{Introduction}
\label{sec:intro}

The understanding of bulk-boundary correspondence in condensed matter systems --- namely, the connection between bulk topological invariants and electronic properties at the system boundary --- has come a long way since the discovery of the quantum Hall effect \cite{PhysRevLett.49.405}. 
This path exhibits as hallmarks the generalization to systems with time-reversal symmetry, which lead to the discovery of three-dimensional (3D) strong topological insulators (TIs) \cite{PhysRevB.75.121306,PhysRevB.79.195322,PhysRevLett.98.106803}, 
the subsequent understanding of the role played by crystalline symmetries, which lead to the identification of weak topological insulators (WTIs) \cite{PhysRevLett.98.106803,ringel2012strong} and of topological crystalline insulators (TCIs) \cite{PhysRevLett.106.106802,hsieh2012topological},
and the comprehension of how the correspondence works in topological semimetals \cite{PhysRevB.83.205101,xu2015discovery,lv2015experimental,soluyanov2015type}.
Recently, the relation between bulk and boundary has been further extended by the discovery of higher-order topological insulators (HOTIs) ~\cite{ Benalcazar2017, Benalcazar2017a, Langbehn2017, Hayashi2018, Song2017, schindler2018higher2, schindler2018higher, Wang2018, Ezawa2018, Ezawa2018a, Ezawa2018d, Dwivedi2018a, Miert2018, Ezawa2018b, Hsu2018, Yan2018, Wang2018a, Trifunovic2018, Geier2018, Liu2018, Serra-Garcia2018, Serra-Garcia2018a, Peterson2018, Zhang2018, Imhof2018}.

The identification of materials that can realize topologically nontrivial phases is naturally important.
Recently, it was predicted that a two-dimensional (2D) monolayer of the naturally occurring mineral called Jacutingaite \cite{vymazalova2012jacutingaite}, \jac, can realize the quantum spin Hall insulator (QSHI) state \cite{,PhysRevLett.120.117701}. 
Furthermore, it was argued that the competition between large spin-orbit coupling, associated with Hg and Pt atoms, and inherent lattice instabilities towards the breaking of inversion symmetry lead to a topological state robust at room temperature and switchable by external electric fields.
First experimental results on the QSHI state have been reported \cite{exp2019}, and it was suggested that the monolayer may become superconducting as well \cite{wu2018unconventional}.
In view of these results, it is of interest to study the topological properties of the electronic structure in the three-dimensional compound.

In this work, we first address this issue and show that \jac\, is both a weak topological phase and a mirror-protected topological crystalline phase.
This result places Jacutingaite in the list of materials which can host surface states protected by different, unrelated crystalline symmetries \cite{rusinov2016mirror, avraham2017coexisting, zeugner2018synthesis, schindler2018higher2, eschbach2017bi, Zhou2018}.
Second, we construct minimal tight-binding models that capture the essential topological properties of this dual topological phase in materials with three-fold rotation symmetry and compare their associated surface spectral density with those of different materials in the class.

The rest of this work is organized as follows. In Sections \ref{sec:topo} and \ref{sec:surf} we use density functional theory to determine the bulk bandstructure of \jac, together with its topological invariants and surface states. We then introduce tight-binding models which reproduce the topology of this material in Section \ref{sec:tb}, and conclude in Section \ref{sec:conc}. The Appendix is devoted to providing results on \jacpd, as well as more details on the tight-binding models.

\section{Topological characterization}
\label{sec:topo}

\begin{figure}[tb]\center
\includegraphics[width=3.7 cm,angle=0,keepaspectratio=true]{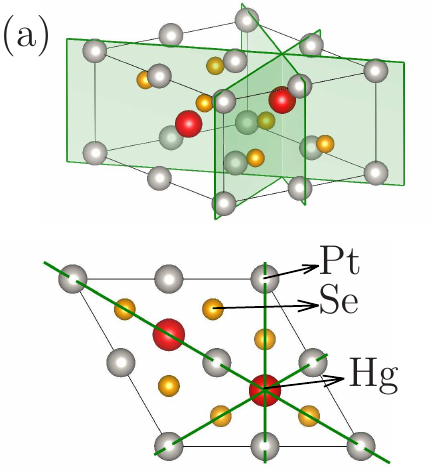}
\includegraphics[width=2.70 cm,angle=0,keepaspectratio=true]{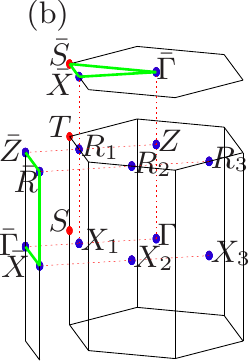}
\vspace{2mm}

\includegraphics[width=8.0 cm,angle=0,keepaspectratio=true]{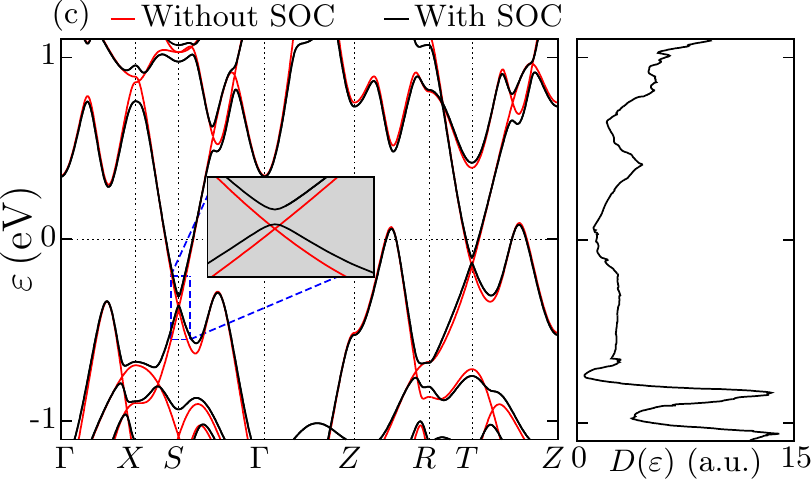}
\caption{Panel (a): Crystal structure of \jac. The three mirror planes are shown as shaded areas (top) and as green lines (bottom). Panel (b): Bulk and surface Brillouin zones (BZ). Time-reversal invariant points are colored blue and points belonging to the surface BZ are indicated with an over-line. Panel (c): Bandstructure and density of states, $D(\varepsilon),$ of \jac\, with (black) and without (red) spin-orbit coupling.\label{bands_plot}}
\end{figure}

\jac\, and \jacpd\, are layered compounds with space group $P\bar{3}m1$ (164) \cite{vymazalova2012jacutingaite,laufek2017powder}. 
The Hg atoms are positioned in a triangular environment delimited by planes of Se and Pt (or Pd) and form a buckled honeycomb lattice [see Fig.~\ref{bands_plot}a].
We performed fully relativistic density functional theory (DFT) calculations using the generalized gradient approximation (GGA) for the exchange and correlation functional \cite{koepernik1999full},\footnote{We used the \textsc{fplo} code \cite{koepernik1999full} version 18.55 with a tetrahedron method for numerical integrations using $16^3$ k-point mesh in the BZ.}. In the following, we base our discussion on the Pt compound, but similar results are obtained for the Pd case (presented in Appendix \ref{app:pd}).
Figure \ref{bands_plot}c shows the bandstructure of \jac~and its density of states, $D(\varepsilon)$, which indicates that the material is a semimetal with small electron and hole pockets.
Similar to Bi \cite{schindler2018higher}, the spin-orbit coupling (SOC) opens a topological gap ($\sim 20\,$meV) 
throughout the full Brillouin zone (BZ) between the bands that gives rise to the electron
pockets and those that host the hole pockets. This allows us to characterize the topological structure of the latter set of bands. 

We first focus on the $\mathbb{Z}_2$ time-reversal polarization invariants \cite{PhysRevB.74.195312,PhysRevLett.98.106803}. 
Since the system has inversion symmetry, we use the Fu-Kane formulas based on the parity of the occupied Bloch states at the time-reversal invariant momenta (TRIM) in the Brillouin zone (BZ) \cite{PhysRevB.76.045302}.
The TRIM are sketched in Fig.~\ref{bands_plot}b and the parity invariants take the values $\delta(X)=\delta(R)=1$ and  $\delta(\Gamma) = \delta(Z) = -1$. 
Hence, we obtain the four $\mathbb{Z}_2$ indexes $(\nu_0;\nu_1\nu_2\nu_3)=(0;001)$, where $\nu_0$ is the strong topological invariant and $\nu_{1,2,3}$ are the three weak topological invariants.
This result is consistent with the idea of engineering weak topological insulators by stacking QSHI layers along one direction, previously realized in KHgSb \cite{PhysRevLett.109.116406}, Bi$_x$Se$_y$ \cite{majhi2017emergence} and Bi$_{14}$Rh$_3$I$_9$ \cite{rasche2013stacked}.
Namely, if the stacking preserves translational invariance along the stacking direction, $z$ in our case, the helical edge states associated with each QSHI layer form surface Dirac cones at momenta $k_z=0,\pi$. As long as translation symmetry is preserved, the surface Dirac cones cannot hybridize and gap out, since they occur at different $k_z$.
Note, however, that not all materials which realize a QSHI in the single layer limit are in a 3D weak topological phase, a counter-example being elemental Bi \cite{Reis2017}. As such the topological invariants of the bulk compound cannot, in general, be inferred from the behavior of a monolayer.

We also note in passing an ambiguity associated with the methodology used for computing the parity invariants. While our calculation includes all of the occupied Kramers pairs, ab-initio methods based on pseudo-potentials necessarily constrain the calculation to states in a certain energy windows and this may result in a global sign difference of the parity invariants. For instance, when performing calculations using the \textsc{vasp} code \cite{PhysRevB.59.1758,hafner2008ab}, we obtained the opposite indexes, +1 at $\Gamma$ and $Z$ and -1 at the other TRIMs \footnote{In the \textsc{vasp} calculation the parity invariants are computed for states in the energy windows $[-17.5,0]$\,eV.}.
This ambiguity is nevertheless of little importance, since predictions regarding the surface spectral density involve in general a product of an even number of parity invariants \cite{teo2008surface}.

An alternative view of the nontrivial topology is provided by the framework of elementary band representations (EBRs) \cite{bradlyn2017topological,vergniory2019complete}. According to this view, the nontrivial topology should be reflected in the fact that it is not possible to decompose the occupied bands into physical EBRs.
In centrosymmetric systems, a distinctive property of physical EBRs is that the number of parity eigenvalues equal to $-1$ per Kramers pair must be a multiple of 4, a fact that provides a simple criterion to assess whether a material is topologically trivial or not \cite{wang2018higher}. 
We find that the number of negative parity eigenvalues modulo 4 to be 2, thus demonstrating nontrivial topology \footnote{We find 64 parity eigenvalues equal to $-1$ at $X$ and $R$ and 61 at $\Gamma$ and $Z$.}.

The stacking of 2D \jac\, layers in 3D Jacutingaite preserves symmetries other than translation, which can lead to additional topological invariants.
In particular, in this work we focus on the reflection symmetries: one mirror plane in momentum space contains the points $\Gamma$, $Z$ and $X$,
while two additional mirrors are obtained by $2\pi/3$ rotations with respect to the $z$ axis [see Fig.~\ref{bands_plot}a and b].
Since a mirror rotates the electron spin by an angle $\pi$, it leads to a wavefunction of opposite sign when applied twice.
Therefore, the mirror symmetry operator satisfies $M^2=-1$ and, 
accordingly, Bloch states belonging to a mirror plane can be labeled with a mirror eigenvalue equal to $\pm i$.
This introduces a partition of the space formed by Bloch states with momentum in the mirror plane into two subspaces and allows to define a Chern number in each of them as
\begin{equation}\label{eq:Cpm}
	C^{\pm} = \frac{1}{2\pi} \sum_n \int_{\mathcal{M}} \mathbf{dS} \cdot \mathbf{\Omega}^{\pm}_n(\mathbf{k}).
\end{equation}
Here, $n$ labels occupied states,  $\mathbf{dS}$ is a differential surface element of the mirror plane $\mathcal{M}$ and $\mathbf{\Omega}^{\pm}_n(\mathbf{k})$ is the Berry curvature associated with the subspace of mirror eigenvalue $\pm i$. It is computed as $\mathbf{\Omega}^{\pm}_n(\mathbf{k})=\mathbf{\nabla}_\mathbf{k}\times\mathbf{A}^{\pm}_n(\mathbf{k})$, with $\mathbf{A}^{\pm}_n(\mathbf{k})=-\mathrm{i}\langle u_\kn^{\pm} | \partial_\mathbf{k}  u_\kn^{\pm} \rangle$ the Berry connection, and $|u_\kn^{\pm}\rangle$ the Hamiltonian eigenstates within a given subspace.
A nonzero mirror Chern number, $C_\mathcal{M} = (C^+-C^-)/2$, signals the existence of topologically protected states on surfaces that preserve the mirror symmetry. 
In \jac\, and \jacpd\, we obtain $C_\mathcal{M}=-2$ for each of the three mirror planes \footnote{As a benchmark, we have computed the mirror Chern number for Bi$_2$Te$_3$, obtaining as a result -1, and for SnTe, obtaining as a result -2, values that agree with the literature. }, which means that these materials can realize helical HOTI phases under suitable surface perturbations, as shown in Ref.~\cite{schindler2018higher2}. Together with the nontrivial weak topological invariant, the nonzero mirror Chern numbers enable us to predict that \jac~is in a dual topological phase, one which is present in a naturally ocurring mineral.

\section{Surface Dirac cones}
\label{sec:surf}

We now present surface spectral densities computed for semi-infinite systems obtained by cutting the crystal perpendicular to the [100] or [001] directions.
In both cases, we chose surfaces terminated by Hg atoms.
For these calculations, we built a tight-binding Hamiltonian based on Wannier functions associated with orbitals 6$s$, 6$p$, and 5$d$ of Hg and of Pt, and 4$p$ of Se \footnote{Se Supplemental Material at [URL TO BE INSERTED] where we provide different input files used for producing the data of this article. These include input files for the FPLO code, as well as input files to generate the definitions of Wannier functions and scripts used for the calculation of the surface spectral densities.}. The Wannier-interpolated band-structure accurately reproduces the DFT results in the energy range $[-9\,\text{eV},5\,\text{eV}]$.

\begin{figure}[tb]\center
\includegraphics[width=8.5 cm,angle=0,keepaspectratio=true]{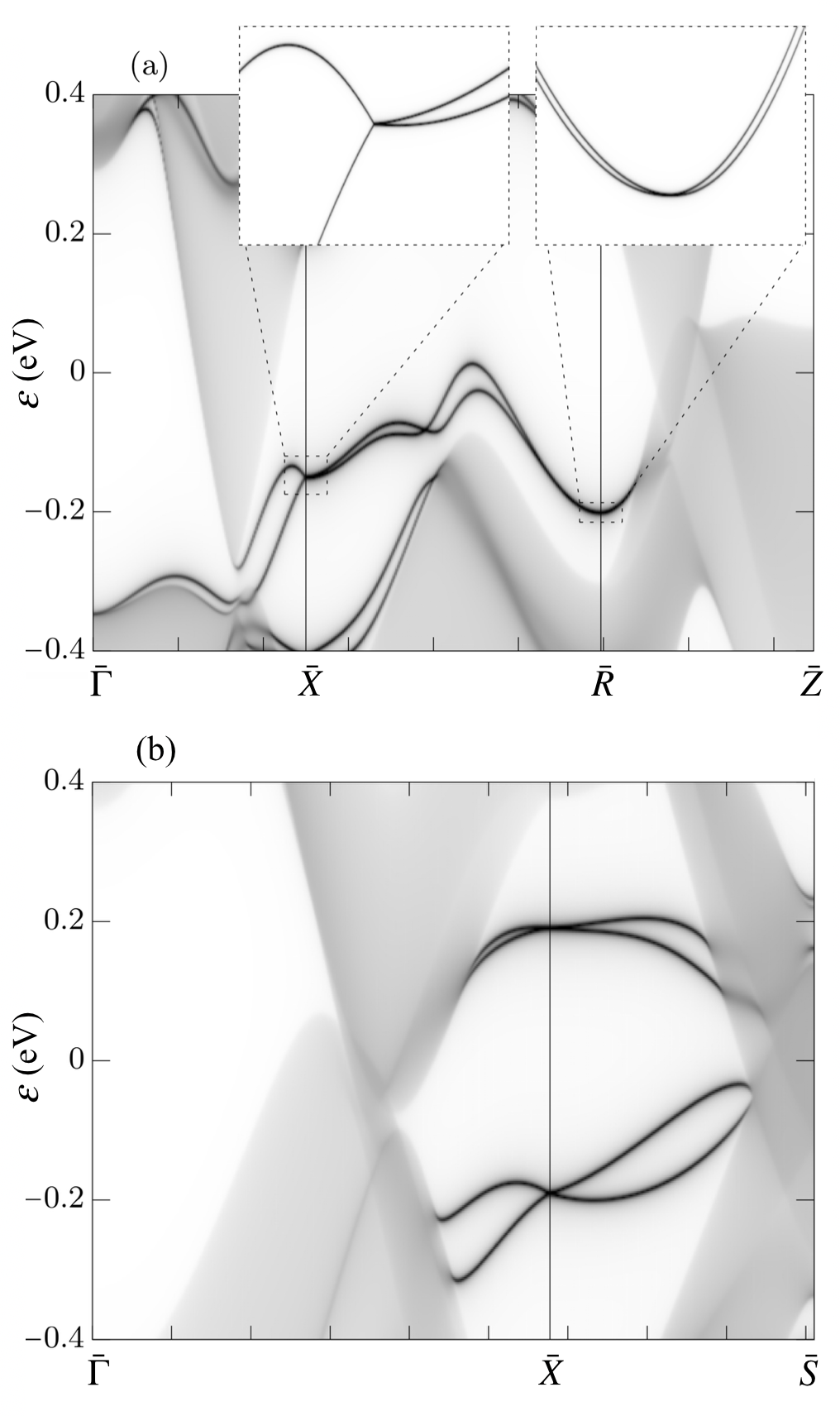}
\caption{Momentum-resolved surface spectral densities. Panel (a): [100] surface. Dirac cones associated with the weak topological index are observed at $\bar{X}$ and at $\bar{R}$. The insets show a zoom in on the Dirac cones.
Panel (b): [001] surface. A pair of Dirac cones associated with the mirror Chern number are observed at $\bar{X}$.\label{surf}
}
\end{figure}

Figure \ref{surf}a shows the momentum-resolved spectral density associated with the [100] surface along a path connecting the surface TRIMs $\bar{\Gamma}$, $\bar{X}$, $\bar{R}$, and $\bar{Z}$ [see Fig.~\ref{bands_plot}b].
The spectral density features two Dirac cones, one centered at $\bar{X}$ and one at $\bar{R}$.
The connectivity of these surface states with the projection of the bulk valence and conduction bands
is consistent with the surface fermion parity invariants, which can be computed from the bulk parity invariants as $\pi(\Lambda_a) = (-1)^{n_b}\delta(\Gamma_i)\delta(\Gamma_j)$ \cite{teo2008surface}. Here, $\Gamma_{i,j}$ are the bulk TRIMs, whose projection coincides with the surface TRIM $\Lambda_a$, and $n_b$ is the number of occupied Kramers pairs.
The opposite parity invariants at $\Gamma$ and $X$ makes $\pi(\bar{\Gamma})$ and $\pi(\bar{X})$ have opposite signs. This reflects a change in the time-reversal polarization on going from $\bar{\Gamma}$ to $\bar{X}$ and, accordingly, along a path connecting these points, surface bands must intersect an odd number of times a generic Fermi energy lying within the bulk topological gap, as observed. On the other hand, $\pi(\bar{X})=\pi(\bar{R})$, such that surface states along this path do not connect bulk valence and conducting states. Lastly, the odd number of intersections should occur also along $\bar{R}$ - $\bar{Z}$, but on this path the bulk state projections overlap, such that the surface gap closes.

Finally, Fig.~\ref{surf}b shows the spectral density associated with the [001] surface. While the projected bulk valence and conducting bands leads to a closing of the surface gap, a pair of Dirac cones are visible at $\bar{X}$, consistent with having $|C_\mathcal{M}|=-2$ \cite{Note2}. 

\section{Tight-binding models}
\label{sec:tb}

\begin{figure}[tb]\center
 \includegraphics[width=0.95\columnwidth]{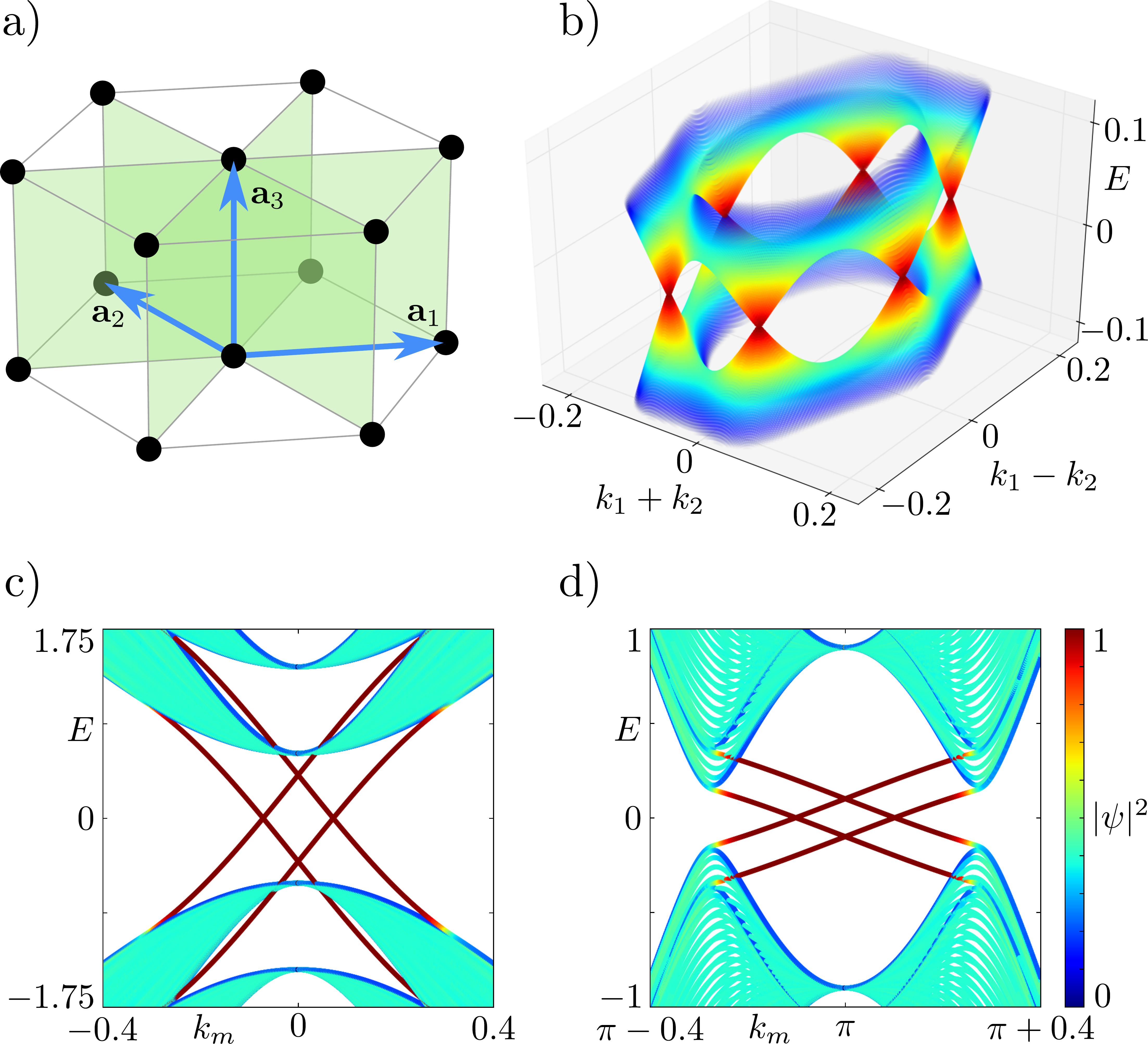}
	\caption{Panel (a): Triangular lattice model of $H_{\rm TI}$, Eqs.~\eqref{eq:HTB} and \eqref{eq:fG}. Sites are shown in black, hoppings in gray, and mirror planes as shaded areas. The blue arrows represent Bravais vectors ${\bf a}_{1,2,3}$. Panel (b): Bandstructure of the model defined by Eqs.~\eqref{eq:fullH} and \eqref{eq:fG} computed in a slab geometry. Only states on the top surface are shown. Panel (c): Cut of the same bandstructure along the $k_1=k_2$ mirror plane, with $k_m$ labeling the momentum along the mirror plane. Panel (d): Bandstructure for model defined by Eqs.~\eqref{eq:fullH} and \eqref{eq:fX}, along the same $k_1=k_2$ mirror plane. All bandstructures are obtained in a slab geometry with hard wall boundary conditions perpendicular to ${\bf a}_3$ and a thickness of 40 unit cells. In (b) and (c) we set $\mu=3$, $\lambda=1$, $\alpha=5$, and $\varepsilon=0.4$, whereas $\varepsilon=0.1$ in panel (d). The color scale in panels (c) and (d) denotes the integrated probability density of wavefunctions on the top- and bottom-most 8 sites of the slab, such that surface modes appear in red (dark gray) and bulk modes in light green (light gray).
\label{fig:tbmodel}
}
\end{figure}

Having established that \jac\, and \jacpd\ simultaneously realize weak and topological crystalline phases,
we will now introduce tight-binding models that capture the essential topological features of this class of materials, which has been found to also include Bi$_2$TeI \cite{rusinov2016mirror,avraham2017coexisting}, Bi$_2$TeBr \cite{zeugner2018synthesis}, BiTe  \cite{eschbach2017bi} and BiSe \cite{schindler2018higher2}.
Specifically, we will construct systems with three-fold rotation symmetry, which are both WTI and TCI 
\footnote{Strictly speaking, Bi$_2$TeI has a small distortion that reduces the symmetry from trigonal to monoclinic, but this distortion is usually neglected, see Ref. \cite{avraham2017coexisting} }.
Our models consist of two coupled strong 3D TIs, a construction similar to the so called ``double strong TI'' of Ref.~\cite{Khalaf2018}. Each strong TI is defined on a triangular lattice with Bravais vectors ${\bf a}_1 = (1, 0, 0)$, ${\bf a}_2 = (-1/2, \sqrt{3}/2, 0)$, and ${\bf a}_3=(0,0,1)$, as shown in Fig.~\ref{fig:tbmodel}a. Its momentum space Hamiltonian takes the form:
\begin{equation}\label{eq:HTB}
\begin{split}
	H_{\rm TI}({\bf k}) = & \Gamma_1 [ \mu +f(k_1,k_2) - \cos(k_3)]  \\
 & +\lambda^\pd [ \Gamma_2 \sin(k_1) + \Gamma_{2,1} \sin(k_2) \\
 & \qquad - \Gamma_{2,2} \sin(k_1+k_2) + \Gamma_3 \sin(k_3) ].
\end{split}
\end{equation}
Here, ${\bf k} = (k_1, k_2, k_3)$ is the crystal momentum vector, where $k_j = {\bf k}\cdot {\bf a}_j$. The matrices $\Gamma_1=\tau_z\sigma_0$, $\Gamma_2=\tau_x\sigma_x$, and $\Gamma_3=\tau_y\sigma_0$, where Pauli matrices $\tau$ encode the degree of freedom associated with two orbitals per site, whereas Pauli matrices $\sigma$ parametrize the spin degree of freedom. Further, the matrices $\Gamma_{2,1}$ and $\Gamma_{2,2}$ are obtained from $\Gamma_2$ by applying a three-fold rotation around the $z$ axis:
\begin{equation}\label{eq:rotGamma}
\Gamma_{2,j} = C_3^j \Gamma_2 C_3^{-j}, \quad C_3=\tau_0 \exp \left( i\frac{\pi}{3}\sigma_z \right).
\end{equation}
The scalar function $f(k_1,k_2)$ encodes the in-plane hoppings of the model, and for Eq. \ref{eq:HTB} to describe a strong TI, it must be such that the system presents band inversions at an odd number of TRIMs. In the following we consider two specific examples of $f(k_1,k_2)$ that will allow us no only to describe the topology of Jacutingaite but also to connect with other materials predicted to be both WTI and TCI \cite{rusinov2016mirror, eschbach2017bi}. The function is such that $H_{\rm TI}$ has either one band inversion at ${\bf k}=(0,0,0)$, or a total of three band inversions at $(0,\pi, 0)$, $(\pi, 0, 0)$, and $(\pi,\pi, 0)$. The resulting functions are:
\begin{align}
	f^{(\Gamma)} =& -\sum_{j=0}^{2} \cos({\cal C}_3^j k_1) \label{eq:fG},\\
	f^{(X)} =& \frac{3}{8} \sum_{j=0}^{2} \cos(2 {\cal C}_3^j k_1)[\cos({\cal C}_3^{j+1} k_1)+\cos({\cal C}_3^{j+2} k_1)-2], \label{eq:fX}
\end{align}
where ${\cal C}_3$ denotes the action of a three-fold rotation on a momentum component: ${\cal C}_3 k_1 = k_2$ and ${\cal C}_3 k_2 = -k_1-k_2$.

For both choices of $f(k_1,k_2)$, the Hamiltonian Eq.~\eqref{eq:HTB} obeys an inversion symmetry $I=\tau_z\sigma_0$, time-reversal symmetry $T=i\tau_0\sigma_yK$ with $K$ complex conjugation, as well as the three-fold rotation symmetry $C_3$ of Eq.~\eqref{eq:rotGamma}:
\begin{equation}\label{eq:HTBC3}
 C_3^\dag H_{\rm TI}(k_1,k_2,k_3) C_3^\pd = H_{\rm TI}(k_2,-k_1-k_2, k_3).
\end{equation}
Its topological structure is determined by the values of $\mu$ and $\lambda$. For both $f^{(\Gamma)}$ and $f^{(X)}$, setting $\mu=3$ and $\lambda=1$, the model realizes a strong 3D TI with $\mathbb{Z}_2$ indices $(\nu_0;\nu_1\nu_2\nu_3)=(1;000)$.

Crucially, the model Eq.~\eqref{eq:HTB} is also mirror symmetric. There exists a mirror symmetry on the $k_1=-2k_2$ plane of the BZ, $M_1 = i\tau_0\sigma_x$,
\begin{equation}\label{eq:HTBM1}
 M_1^\dag H_{\rm TI}(k_1,k_2,k_3) M_1^\pd = H_{\rm TI}(k_1,-k_1-k_2,k_3),
\end{equation}
as well as two other mirror symmetries obtained through rotation, $M_2 = C_3^{-1} M_1 C_3^\pd$ and $M_3 = C_3^{-2} M_1 C_3^2$, with mirror planes 
$k_2=-2k_1$ and $k_1=k_2$, respectively. On each mirror-invariant plane, $H_{\rm TI}$ can be block-diagonalized into sectors corresponding to mirror eigenvalues $\pm i$. By computing the Chern number associated with each sector, Eq.~\eqref{eq:Cpm}, we find a mirror Chern number $C_{\cal M}=-1$ on each of the mirror planes. As such the tight-binding model of Eq.~\eqref{eq:HTB} is simultaneously a strong 3D TI and a TCI.

We will follow our construction of a dual WTI and TCI Hamiltonian using the case $f = f^{(\Gamma)}(k_1,k_2)$, for which the Hamiltonian takes the form used in Ref.~\cite{schindler2018higher} and hosts a single Dirac cone on each surface of the system, positioned at the $\bar{\Gamma}$ point of the surface BZ. Similar steps are presented in Apps. \ref{app:tb} and \ref{app:realspace} for the case defined by Eq.~\eqref{eq:fX}.
We consider two copies of $H_{\rm TI}$ which are displaced relative to each other by half of a unit cell in the $z$ direction \cite{Queiroz2018}. The full momentum space Hamiltonian is an $8\times 8$ matrix having the block form
\begin{equation}\label{eq:fullH}
H({\bf k}) = \begin{pmatrix}
    H_{\rm TI}(k_1,k_2,k_3)+\varepsilon & \alpha A(k_1,k_2) \\
    \alpha A(k_1,k_2) & H_{\rm TI}(k_1,k_2,k_3+\pi)-\varepsilon \\
             \end{pmatrix},
\end{equation}
where $\varepsilon$ is a relative energy shift between the two TI blocks, $A(k_1,k_2)$ is a coupling term (to be defined later), and $\alpha$ is its strength.

For $\alpha=0$, the Hamiltonian Eq.~\eqref{eq:fullH} is a block-diagonal double strong TI which obeys the same symmetries as $H_{\rm TI}$. Time-reversal, inversion, rotation, as well as the three mirror symmetries have the same matrix structure in $\tau$ and $\sigma$ space, and are block-diagonal in the space of the two 3D TIs. There are now, however, two band inversions in total. The upper block contributes a band inversion at ${\bf k}=0$, as discussed before, whereas the lower one has inverted bands at ${\bf k}=(0,0,\pi)$, due to the extra momentum shift. As such, the parity invariants associated with Eq.~\eqref{eq:fullH} are identical to \jac, $\delta(\Gamma)=\delta(Z)=-1$, marking it as a WTI with $\mathbb{Z}_2$ indices $(\nu_0;\nu_1\nu_2\nu_3)=(0;001)$. A [100] surface will exhibit two surface Dirac cones, one at $k_z=0$ and one at $k_z=\pi$, protected by time-reversal and translation along $z$.

By combining two TI blocks, we have obtained a system with a trivial strong index, $\nu_0=0$, due to the latter's $\mathbb{Z}_2$ classification. Mirror Chern numbers, on the other hand, have an integer classification, which means that the Hamiltonian Eq.~\eqref{eq:fullH} is also a TCI with mirror Chern numbers given by the sum of the invariants in each block. We find $C_{\cal M}=-2$ on each of the three mirror planes, reproducing the behavior of \jac\, and \jacpd. 
Due to this dual topology, $H$ will exhibit surface Dirac cones not only on its side surfaces, but on the top, [001] surface as well. For $\alpha=0$, there are two surface Dirac cones positioned at the $\bar{\Gamma}$ point of the surface BZ and at energies $\pm\varepsilon$. 

As long as the two TI blocks remain uncoupled, the surface Dirac cones are orthogonal to each other, such that a circular band crossing occurs at $E=0$ whenever $\varepsilon\neq0$. This band crossing, however, is an artifact of the block-diagonal nature associated with the choice $\alpha=0$. For $\alpha\neq0$, the two TI blocks couple, lifting the degeneracy. We choose an off-diagonal coupling term
$A(k_1,k_2) = \tau_x\sigma_z\sum_{j=0}^{2} \sin\big({\cal C}_3^j (k_1-k_2)\big)$ 
which preserves time-reversal, inversion, as well as rotation and mirror symmetries. Due to this term, the circular band crossing is lifted everywhere except on the three mirror planes. The result is that the [001] surface now hosts a total of six Dirac cones. 
On each of the three mirror planes there exists a pair of Dirac cones positioned symmetrically around $\bar{\Gamma}$ due to time-reversal symmetry, as shown in Fig.~\ref{fig:tbmodel}b for a slab geometry calculation. Their position can be tuned with the relative energy shift of the two TI blocks, $\varepsilon$. 

Fig.~\ref{fig:tbmodel}c shows the bandstructure along a mirror plane in a larger energy scale. Notice that the branches of the two $E=0$ Dirac cones positioned symmetrically around $\bar{\Gamma}$ can intersect at the TRIM giving rise to a second pair of Dirac cones. In a given material, details of the bulk bandstructure and of the surface potential will dictate if all or some of these four Dirac cones are observed.
For instance, DFT calculations in Bi$_2$TeI \cite{rusinov2016mirror} have highlighted Dirac cones both at $\bar{\Gamma}$ and 
away from the TRIM, and similar results are found in Bi$_2$TeBr \cite{zeugner2018synthesis} and in BiTe \cite{eschbach2017bi}. 
Different to these materials, Jacutingaite has two observable surface Dirac cones at $\bar{X}$ (see Fig.~\ref{surf}b), while the possible Dirac cones away of the TRIMs are hidden by the projection of the bulk spectral density.
This scenario is better described by Hamiltonian Eq. \ref{eq:fullH} with the choice $f(k_1,k_2) = f^{(X)}$. In this case, as
Fig.~\ref{fig:tbmodel}d depicts, the Dirac cones are positioned closer to and symmetrically with respect to $(\pi,\pi,0)$.

\section{Conclusion}
\label{sec:conc}

We have shown that the naturally ocurring and recently synthesized \jac~and \jacpd~belong to the class of systems that simultaneously realize weak and topological crystalline phases.
In addition, we have introduced a set of tight-binding models that contain the essential properties of this dual topological phase, reproducing the main features of the surface spectral densities which have been predicted for different materials in the class. 

We note that, similar to elemental bismuth \cite{schindler2018higher}, these topologically non-trivial materials are not insulators, but semimetals with small electron and hole pockets. Since they lack a bulk mobility gap, it is unlikely that they will show surface dominated transport, due to the fact that disorder, which is unavoidable in any realistic setup, will scatter surface electrons into the bulk. However, the systems we have studied have well defined topological gaps throughout their BZ, meaning that weak and strong $\mathbb{Z}_2$ topological indices as well as mirror Chern numbers can be meaningfully computed. These invariants necessarily lead to
surface states in the energy range defining the topological gap, shown in Section \ref{sec:surf}, which may be visualized using energy and/or momentum sensitive techniques, such as angle-resolved photo-emission spectroscopy, or scanning tunneling spectroscopy.
Thus, we hope that our work will motivate experimental studies on Jacutingaite, aimed at probing the topology of its surface modes.

Further, we hope that the tight-binding models we have introduced will allow for a better theoretical understanding of WTI+TCI materials. These include the behavior of the system under various surface perturbations, for which it can become a HOTI \cite{schindler2018higher2}, the behavior of modes localized to step edges, which have recently been reported in Bi$_2$TeI \cite{avraham2017coexisting}, as well as the effect of disorder. Moreover, they may be used to understand the degree to which surface modes can influence transport properties in the presence of coexisting bulk states, similar to the studies done for 2D systems in Refs.~\cite{Baum2015, Baum2015a}.

\acknowledgments

We thank Ulrike Nitzsche for technical assistance and acknowledge support from the German Research Foundation (DFG) via SFB 1143, project A5 and the Wurzburg-Dresden Excellence initiative. J.I.F. thanks the IFW excellence programme.

\bibliography{ref}

\clearpage
\newpage

\appendix

\section{Ab-initio results for Pd$_2$HgSe$_3$}
\label{app:pd}

\begin{figure}[t]\center
\includegraphics[width=8cm]{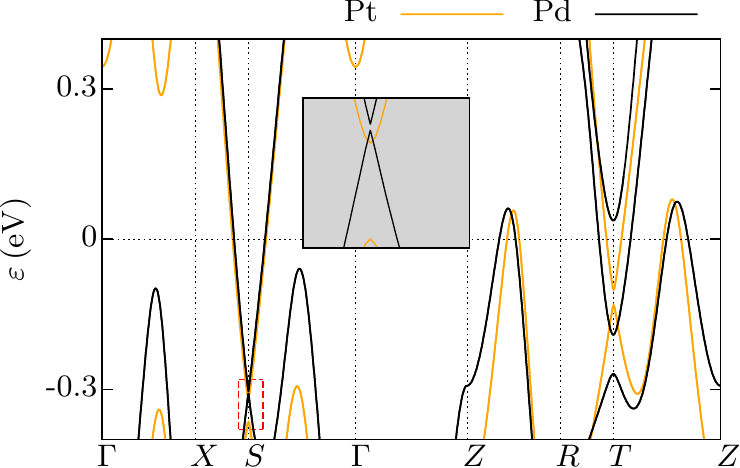}
\caption{ Bandstructure of \jacpd and of \jac. The inset presents a zoom of the area enclosed by the red dotted-line square.\label{fig:bandsPd}}
\end{figure}

We performed Density Functional Theory (DFT) calculations for \jacpd{}
using the crystal structure reported in Ref. \cite{laufek2017powder}.
We used the same calculation setup as in the main text of this article.
The energy dispersion of \jacpd{} and of \jac{} present the same features, as shown in Fig. \ref{fig:bandsPd}.
The main difference is the smaller topological gap between valence and conducting states observed in the Pd based compound.
We obtained for \jacpd\,  the same parity invariants as for \jac, and hence the same time reversal polarization invariants.
Namely, $\delta(X)=\delta(R)=1$,  $\delta(\Gamma) = \delta(Z) = -1$ and $\mathbb{Z}_2=(0;001)$.

\begin{figure}[t]\center
\includegraphics[width=8. cm,angle=0,keepaspectratio=true]{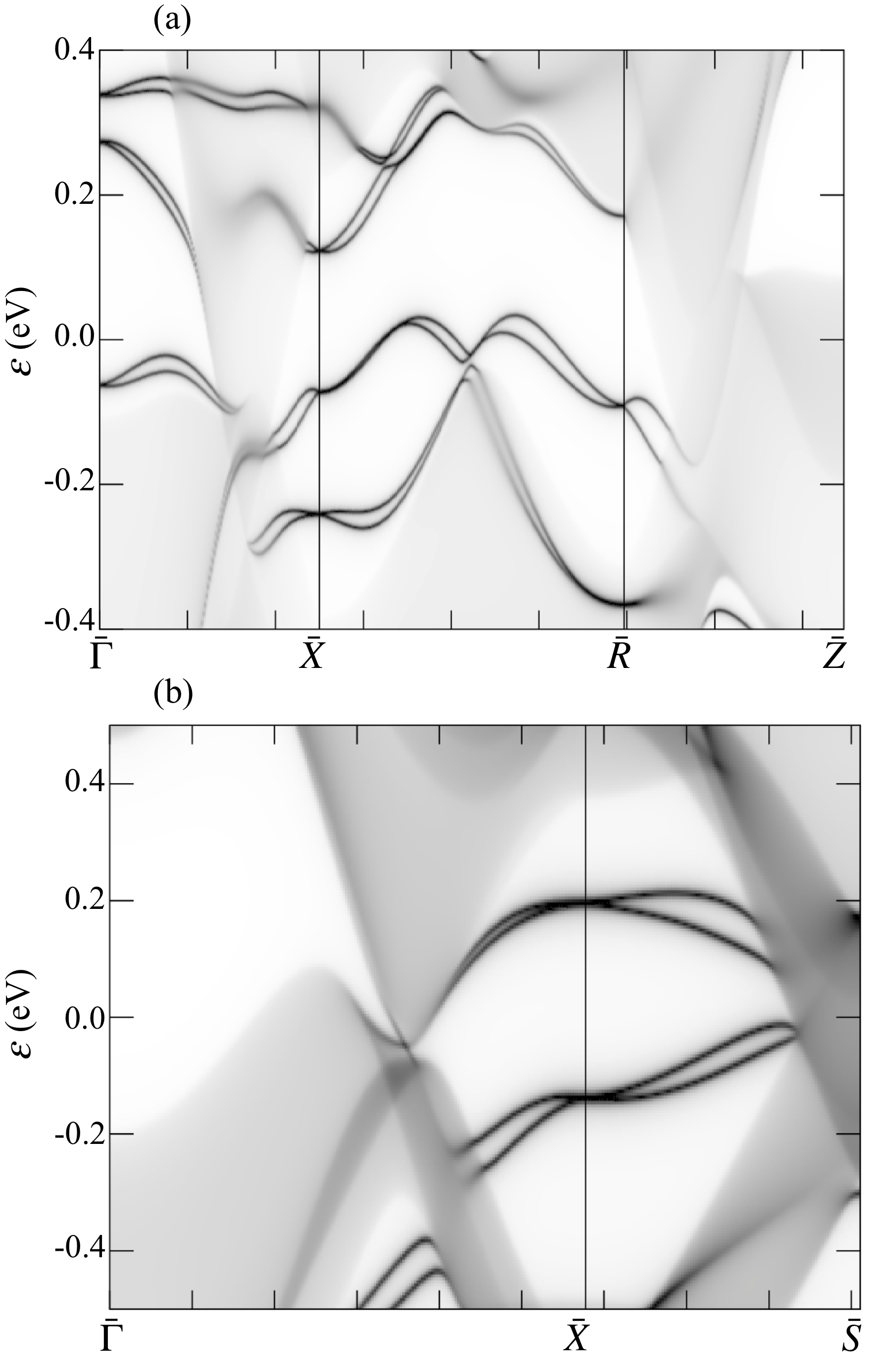}
\caption{Momentum-resolved surface spectral densities of \jacpd. Left: [100] surface. 
Right: [001] surface. A pair of Dirac cones associated with the mirror Chern number are observed at $\overline{X}$.\label{fig:surfbands_app}
}

\end{figure}

We also built a tight-binding Hamiltonian for \jacpd\, based on Wannier functions associated with orbitals 6$s$, 6$p$ and 5$d$ of Hg, 5$s$, 5$p$ and 4$d$ of Pd and 4$p$ of Se and used this Hamiltonian for computing surface spectral densities.
Fig. \ref{fig:surfbands_app}(a) presents the spectral densities at the surface [100]. While surface Dirac cones are visible at $\overline{X}$ and $\overline{R}$, the projection of the bulk bands closes the surface gap between surface time-reversal invariants momenta of different surface parity invariant, and hence, does not allow us to analyze the connectivity between the surface Dirac cones and bulk valence and conduction states. Fig. \ref{fig:surfbands_app}(b) shows the spectral density associated with the surface [001] which, as in \jac, features two Dirac cones at $\overline{X}$.

\section{Different surface Dirac cone positions}
\label{app:tb}

As explained in the main text, the [001] surface spectral density of \jac\, presents two Dirac cones at $\bar{X}$ (and at the other TRIMs connected to $\bar{X}$ by $2\pi/3$ rotations).
Within the 8-band tight-binding models considered in this work, this surface spectral density is better described by the model which presents band inversions at the boundaries of the 3D BZ, such that $\delta(X)=\delta(R)=-1$ and $\delta(\Gamma)=\delta(Z)=+1$. 
These parity invariants differ in a global sign with those computed with \textsc{fplo}, but as the comparison with the \textsc{vasp} calculations shows, they coincide with the ab-initio results when the parity invariant calculation is constrained to the energy windows $[-17.5,0]$ eV.

\begin{figure}[tb]\center
 \includegraphics[width=0.85\columnwidth]{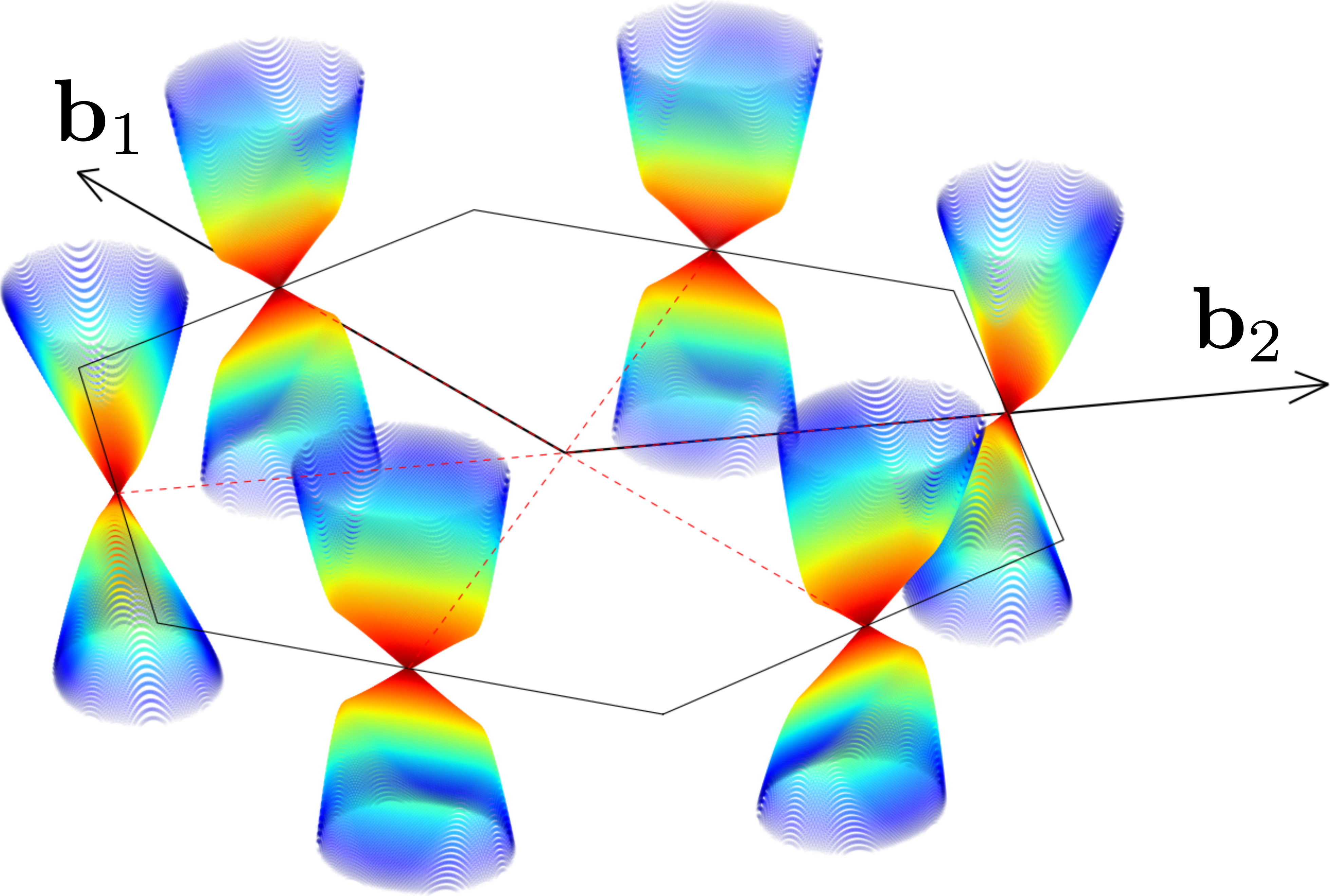}
 \caption{Bandstructure of the model obtained using Eq.~\eqref{eq:fk1k2} in an infinite slab geometry, with thickness of 40 unit cells in the $z$ direction, using $\mu=3$ and $\lambda=1$. Only states which are localized on the top surface and have energies $|E|\leq1$ are shown. To better visualize the surface Dirac cones, we set $\alpha=\varepsilon=0$, such that each Dirac cone is doubly degenerate. The hexagonal contour marks the boundary of the surface BZ, mirror invariant lines are shown in dashed red, and black arrows indicate the reciprocal lattice vectors of the surface ${\bf b}_1=(\sqrt{3}/2, 1/2)$ and ${\bf b}_2=(0, -1)$. \label{fig:surfD2}}
\end{figure}

In the following, we reformulate the construction presented in the main text for this case.
We follow the same strategy, namely to start with two 3D TI blocks. However, unlike $H_{\rm TI}$ of Eq.~(2) in the main text, which had a single band inversion at $\Gamma$, we choose a TI block which has a total of three band inversions, positioned at the three $X$ points of the BZ.
For the TI Hamiltonian presented in the main text, Eq.~\eqref{eq:HTB}, we now focus on the case in which the function $f(k_1,k_2)$ is
\begin{widetext}
\begin{equation}\label{eq:fk1k2}
\begin{split}
	f(k_1,k_2)=f^{(X)}(k_1,k_2) = \frac{3}{8}\Big\{ & \cos(2k_2) \big[ -2 + \cos(k_1) + \cos(k_1 + k_2) \big] \\ 
 & + \cos(2 k_1) \big[ -2 + \cos(k_2) + \cos(k_1 + k_2) \big] \\
 & + \cos(2(k_1 + k_2)) \big[ -2 + \cos(k_1) + \cos(k_2) \big] \Big\}.
\end{split}
\end{equation}
\end{widetext}
In the next section, we detail the real-space structure of these hopping terms. Due to the form of Eq.~\eqref{eq:fk1k2}, $H_{\rm TI}$ obeys the same symmetries as the TI model introduced in the main text: time-reversal, inversion, three-fold rotation, as well as the three mirrors. Setting as before $\mu=3$ and $\lambda=1$, the model realizes a simultaneous strong TI with $\mathbb{Z}_2$ indices $(\nu_0;\nu_1\nu_2\nu_3)=(1;000)$ and TCI with mirror Chern numbers $C_{\cal M}=-1$ on each of the three mirror planes. Crucially however, there are now three band inversions in the model, occurring at $(k_1,k_2,k_3)=(\pi,0,0)$, $(0, \pi,0)$, and $(\pi, \pi,0)$, such that the [001] surface hosts a total of three Dirac cones, positioned at the $\overline{X}$ points of the surface BZ. By forming a double strong TI as in Eq.~\eqref{eq:fullH}, we obtain a dual topological phase, which is simultaneously a WTI and a TCI with $C_{\cal M}=-2$. Figure \ref{fig:surfD2} shows the bandstructure of the model in a slab geometry, infinite in both $k_1$ and $k_2$ and with a thickness of $40$ unit cells in the $z$ direction.

\section{Real-space hopping terms}
\label{app:realspace}

\begin{figure}[tb]\center
 \includegraphics[width=\columnwidth]{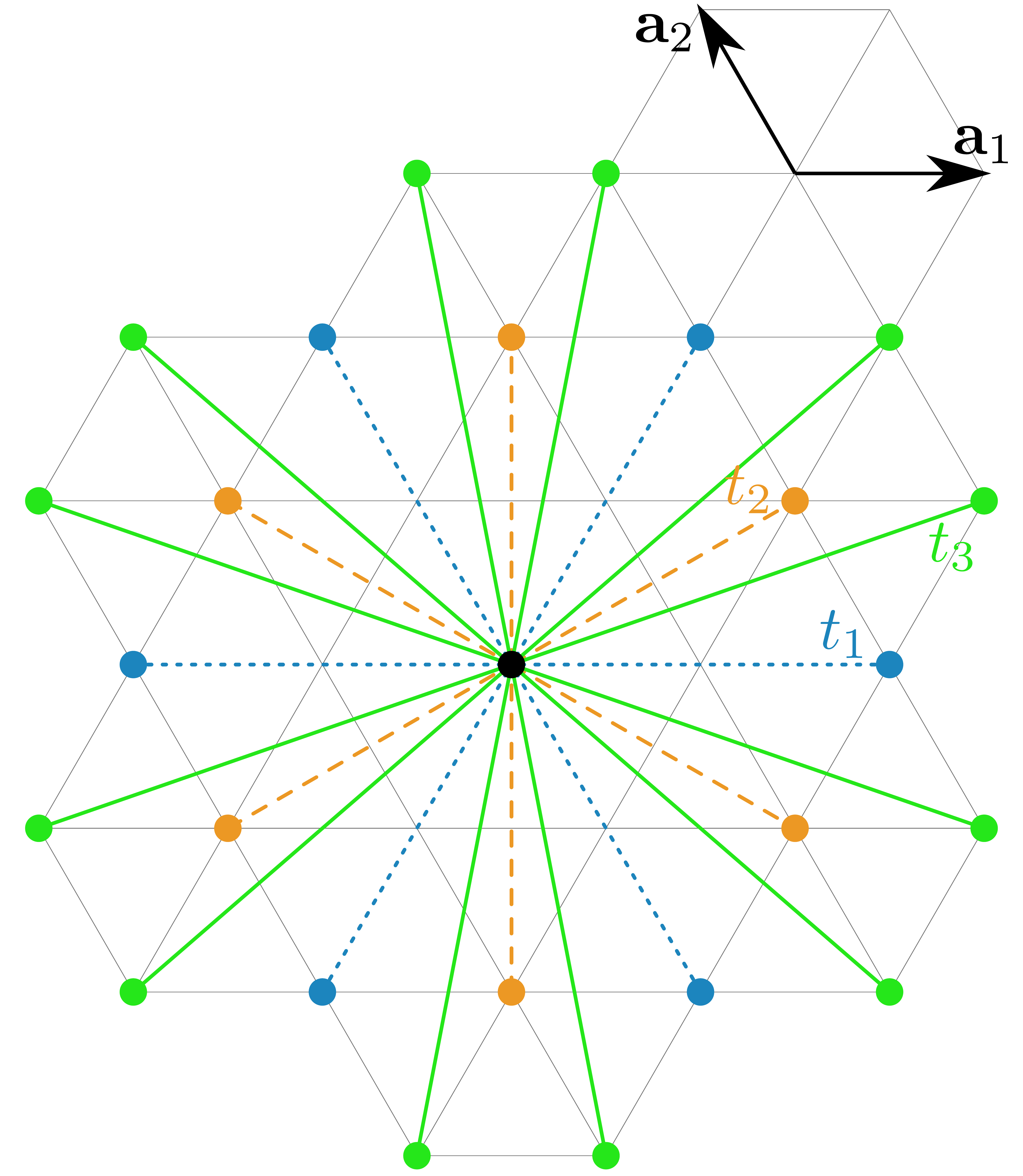}
 \caption{Sketch of the real-space hopping terms corresponding to the function $f^{(X)}$ of Eq.~\eqref{eq:fk1k2}. Shown is the triangular lattice describing $H_{\rm TI}$ at constant $z$ coordinate. Starting from a site on The hopping amplitudes are $t_1=-3/8$ (blue dotted lines), $t_2=3/16$ (orange dashed lines), and $t_3=3/32$ (green solid lines). The structure of the hoppings preserves the three-fold rotation symmetry of the model.\label{fig:hopping}}
\end{figure}

The real-space structure of the Hamiltonian $H_{\rm TI}$ of Eq.~\eqref{eq:HTB} together with the function $f^{(\Gamma)}$ of Eq.~\eqref{eq:fG} are the same as in Ref.~\cite{schindler2018higher2}. For the model with three band inversions at the $X$ points, however, the momentum-space function $f^{(X)}$ of Eq.~\eqref{eq:fk1k2} leads to longer ranged, in-plane hoppings in real space. These hoppings, which all have the same matrix structure, $\Gamma_1=\tau_z\sigma_0$, are shown schematically in Fig.~\ref{fig:hopping}. Their amplitudes are $t_1=-3/8$, $t_2=3/16$, and $t_3=3/32$.

\end{document}